\documentstyle[epsfig,prl,aps,tighten,twocolumn]{revtex}
\newcommand{\be}{\begin{equation}}
\newcommand{\ee}{\end{equation}}
\begin{document}
\draft
\title{\bf Superluminal pulse transmission through a phase-conjugating 
mirror}

\author{M. Blaauboer,$^{\rm a}$  A. E. Kozhekin, $^{\rm b}$  A. G. Kofman, 
$^{\rm b}$ G. Kurizki, $^{\rm b}$  D. Lenstra, $^{\rm a}$  and  
A. Lodder,$^{\rm a}$}

\address{$^a$Faculteit Natuurkunde en Sterrenkunde, Vrije Universiteit,
         De Boelelaan 1081, 1081 HV Amsterdam, The Netherlands \\
$^b$Chemical Physics Department, Weizmann Institute of Science, 
Rehovot 76100, Israel}

\date{\today}
\maketitle

\begin{abstract}
We theoretically analyze wave packet transmission through  
a phase-conjugating mirror and show that the
transmission of a suitably chosen input pulse is
superluminal, i.e. the peak of the pulse emerges from
the mirror before the time it takes to travel the same 
distance in vacuum. This pulse reshaping effect can
be attributed directly to the dispersion relation in the 
nonlinear medium constituting the mirror. Thus, for the first time 
a connection is laid between optical phase conjugation and 
superluminal behavior. In view of its additional amplifying ability, a 
phase-conjugating mirror is a most promising candidate for 
an experimental observation of tachyonic signatures.
\end{abstract}

\pacs{PACS numbers: 42.65.Hw, 42.25.Bs, 42.68.Ay, 73.40.Gk
	           {\tt physics/9711011}}
\narrowtext


It is well known that transmission of a wave packet through an absorbing 
medium gives rise to pulse reshaping by dispersion\cite{chu82}. 
The pulse seems to have been transmitted with a speed larger 
than the speed of light, or "superluminally". Similar behavior 
has recently been observed in tunneling experiments using  
nondissipative dielectric mirrors\cite{spielmann94} and has been explained in 
terms of destructive interference between causally propagating 
consecutive components of the pulse\cite{japha96}. Superluminal particles, or 
"tachyons", were first fully studied in the sixties\cite{terletskii60}.
Theoretically, several models were developed which give rise to 
tachyonic collective modes in, for example, systems of inverted 
pendula\cite{aharonov69} or inverted two-level atoms\cite{chiao96}. However,
experiments carried out to directly observe tachyon-like excitations 
were so far without success\cite{danburg71}. 
Here, we consider a phase-conjugating mirror (PCM) consisting of a pumped 
nonlinear optical material\cite{fisher83}. The dispersion relation 
in this material is shown to be tachyonic, giving rise to group velocities 
larger than the speed of light.
We analyze the transmission of a gaussian wave packet incident 
upon the PCM and find that its peak can be transmitted superluminally. 
This effect is a direct consequence of the dispersion relation and
does not violate causality, since the peak of the transmitted pulse 
is not causally related to the peak of the input, but originates
from the forward tail of this incoming pulse. In order to transmit 
information one could eg. use a discontinuous incident signal. 
This kind of signal also exhibits the superluminal peak-advancement, 
but the discontinuity (information) is transmitted with the speed of 
light, in full agreement with causality. A measurement of pulse 
transmission through a PCM thus provides an experimental signature 
of optical tachyonic excitations.

Our PCM consists of an optical medium with a large 
third-order susceptibility $\chi^{(3)}$. The medium is pumped 
by two intense counterpropagating laser beams of frequency $\omega_{0}$.
When a weak probe beam of frequency $\omega_{0} + \delta$ is incident 
on the material, a fourth beam will be generated due to the nonlinear 
polarization of the medium. This so-called conjugate wave
propagates with frequency $\omega_{0} - \delta$ in the opposite 
direction as the probe beam\cite{fisher83}.

\begin{figure}
\centerline{\epsfig{figure=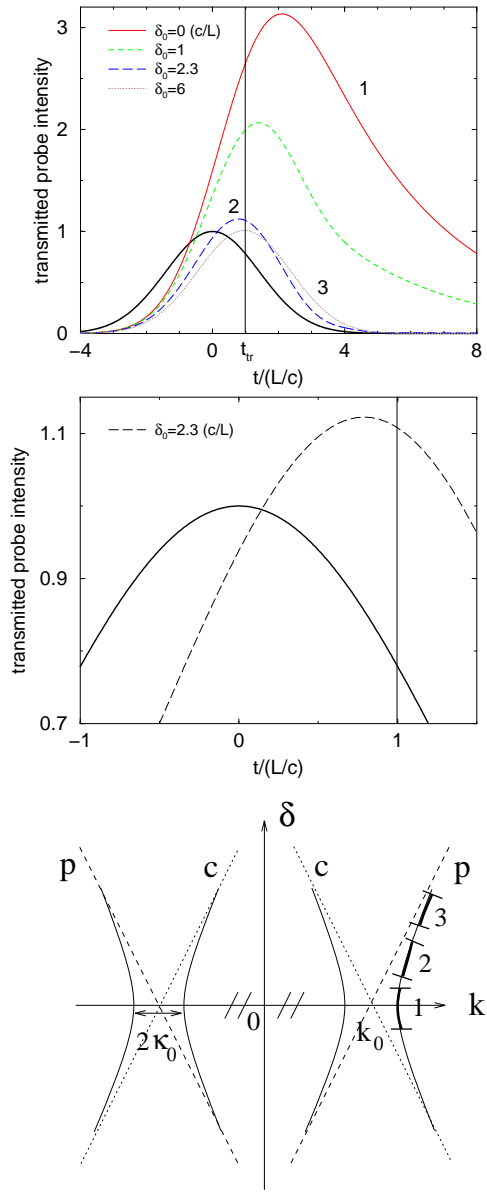,width=0.8\hsize}}
\caption[]{(a) Transmitted probe pulse $|{\cal E}_{p}(L,t)|$ at $x=L$ for an
    incoming gaussian ${\cal E}_{p}(0,t) = e^{-\alpha t^2}
    e^{i \delta_{0} t}$ at $x=0$ (thick solid line). The temporal width
    (FWHM) of the incoming pulse $\Delta_{t}\equiv 
    \frac{2\sqrt{ln2}}{\sqrt{\alpha}}=3.3\, L/c$, the spectral
    width $\Delta_{\delta}=1.7\, c/L$ and $\kappa_{0}L = 1.4$.
    The vertical line indicates the time $t_{tr}=L/c$ needed
    to traverse the cell in vacuum. (b) Enlargement of (a) for 
    $\delta_{0}=2.3\, c/L$, clearly showing advancement of the peak 
    upon transmission. (c) Dispersion relation (solid line) in a 
    $\chi^{(3)}$-material.
The dashed (dotted) lines correspond to the dispersion relation 
for the \underline{p}robe (\underline{c}onjugate) waves in vacuum.
$k_{0} \equiv \omega_{0}/c$. The marked positions 1-3 indicate the 
frequency components in the incident pulses which give rise to the 
transmitted curves 1-3 in (a).}
\label{fig:gauss}
\end{figure}

In order to derive the dispersion relation for an electromagnetic
excitation in this pumped medium, we consider 
the one-dimensional wave equation for a nonmagnetic, nondispersive material
in the presence of a nonlinear polarization
\be
\left( \frac{\partial^2}{\partial x^2} - \epsilon_{r} \epsilon_{0} \mu_{0}
\frac{\partial^2}{\partial t^2} \right) E(x,t) = \frac{1}{\epsilon_{0} c^2}
\frac{\partial^2}{\partial t^2} P_{NL}(x,t)
\label{eq:waveeqn}
\ee
with $P_{NL}(x,t) = \chi^{(3)} E^3(x,t)$. The total field is taken 
as the sum of four monochromatic plane waves
\be
E(x,t) = \sum_{\alpha=1,2,p,c} E_{\alpha} (x,t) = \sum_{\alpha=1,2,p,c}
{\cal E}_{\alpha}(x) e^{-i\omega_{\alpha} t} + c.c.,
\label{eq:totfield}
\ee
where we have labeled the two pump beams as 1 and 2 and the probe and 
conjugate as p and c. ${\cal E}_{\alpha}$ denotes the complex amplitude 
of the $\alpha^{th}$ field which propagates with frequency $\omega_{\alpha}$.
We substitute (\ref{eq:totfield}) in (\ref{eq:waveeqn}) 
and select the phase-conjugation terms in the polarization. 
Assuming the pump beams to be 
non-depleted and applying the appropriate phase-matching 
conditions\cite{fisher83}, we arrive at two coupled equations for 
the amplitudes of the probe and conjugate waves which are, using 
$\delta \ll \omega_{0}$ and taking 
$\epsilon_{r}=1$\cite{lenstra90}
\be
\left( \begin{array}{cc}
- \frac{c^2}{2\omega_{0}} \frac{\partial^2}{\partial x^2} - 
\frac{\omega_{0}}{2}  & - \kappa\, c \vspace{0.3cm} \\
\kappa^{*}\, c &  \frac{c^2}{2\omega_{0}} \frac{\partial^2}{\partial x^2}
+ \frac{\omega_{0}}{2} \end{array} \right)
\left( \begin{array}{l}
{\cal E}_{p}(x) \vspace{0.3cm} \\ {\cal E}_{c}^{*}(x)
\end{array} \right) = \delta
\left( \begin{array}{l}
{\cal E}_{p}(x) \vspace{0.3cm} \\ {\cal E}_{c}^{*}(x)
\end{array} \right).
\label{eq:SEFL}
\ee
Here $\kappa \equiv \kappa_{0} e^{i\phi} = \frac{3\omega_{0}}{\epsilon_{0}c}
\chi^{(3)} {\cal E}_{1} {\cal E}_{2}$ is the pumping induced coupling 
strength (per unit length) between the probe and conjugate wave. 
Because the above matrix is anti-hermitian, the system is 
said to be dissipatively coupled\cite{spreeuw93}.
Trying a harmonic wave solution in (\ref{eq:SEFL}) 
yields the dispersion relation
\be
k^2 = \frac{\omega_{0}^2}{c^2} \pm \frac{2\omega_{0}}{c^2} 
\sqrt{\delta^2 + (\kappa_{0}c)^2}, 
\label{eq:tachdisp}
\ee
which is plotted in Fig.~\ref{fig:gauss}(c). In vacuum $\kappa_{0}=0$, 
and (\ref{eq:tachdisp}) reduces to the four solutions 
$k = \pm (\omega_{0} \pm \delta)/c$.
Note that the group velocity in the nonlinear medium is always
larger than {\it c}, the speed of light in vacuum.

\begin{figure}
\centerline{\epsfig{figure=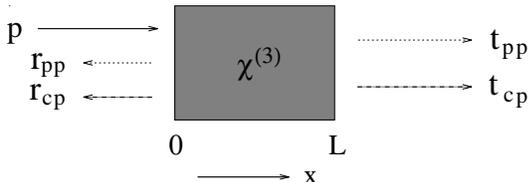,width=0.8\hsize}}
\caption{Reflection and transmission in one dimension of a probe beam 
(solid line) incident from vacuum on a phase-conjugating mirror. 
Dotted (dashed) arrows denote probe (conjugate)
reflected and transmitted beams. 
}
\label{fig:pcm}
\end{figure}

In view of this dispersion relation, 
the question arises how wave packets will be transmitted through a PCM. 
Because of the superluminal group velocity, tachyonic effects are expected. 
We begin by considering the transmission amplitude for a monochromatic probe 
beam incident on a PCM, see Fig.~\ref{fig:pcm}.
Matching the plane wave solutions of (\ref{eq:SEFL}) in the PCM 
at $x=0$ and $x=L$ to the probe and conjugate waves 
outside the cell yields for the probe transmission amplitude 
at $x=L$ the well-known result (choosing $\phi=0$)\cite{fisher83},
\be
t_{pp} (\delta)  =  \frac{\beta} 
{\beta \mbox{\rm cos} (\beta L)  - i\frac{\delta}{c} 
\mbox{\rm sin} (\beta L)}
\label{eq:probetrans}
\ee
where
\be
\beta = \frac{1}{c} \sqrt{\delta^2 + (\kappa_{0} c)^2}.
\ee
Now consider a probe pulse ${\cal E}_{p}(0,t)$ $=$ $\int_{-\infty}^{\infty} 
d\delta\,$ $\tilde{{\cal E}}_{p}(0, \delta)$ $ e^{-i\delta t}$ incident at $x=0$.
Using an analogous two-sided Laplace Transform technique as 
Fisher {\it et al.} \cite{fisher81} employed for phase-conjugate 
reflected pulses,
we obtain for the probe pulse at $x=L$
\be
{\cal E}_{p}(L,t) = \frac{1}{2\pi i} \int_{\gamma - i\infty}^{\gamma
+ i \infty} ds\, t_{pp}(is)\, \tilde{{\cal E}}_{p}(0,s)\, e^{st} 
\label{eq:transpulse}
\ee
with 
\be
\tilde{{\cal E}}_{p}(0,s) = \int_{-\infty}^{\infty} dt^{'} {\cal E}_{p}(0, t^{'})
 e^{-st^{'}}.
\ee
${\cal E}_{p}(L,t)$ is thus expressed as an integral in the 
complex $s$-plane of the input pulse times the transmission amplitude
$t_{pp}$\cite{validity}. The integration contour in (\ref{eq:transpulse})
defined by $\gamma$ must be chosen to the right of all singularities 
of $t_{pp}$\cite{fisher81}.

A direct calculation of these singularities leads to two possible 
operating regimes of the PCM, in which the pulse-reshaping upon transmission
is entirely different. For $\kappa_{0} L < \pi/2$, all singularities of $t_{pp}$
lie in the left half $s$-plane\cite{fisher81} and ${\cal E}_{p}(L,t)$
is always finite. On the other hand, if $\kappa_{0} L > \pi/2$ there will be
at least one singularity in the right half $s$-plane.  
We then find
exponential growth of the transmitted probe field and hence the mirror
is said to be in an unstable (or "active") operating regime\cite{fisher83}. 
Here we restrict ourselves to the case of stable operation. The effects 
of the instability will be discussed elsewhere\cite{blaauboer}.
In view of the above, for $\kappa_{0} L < \pi/2$, we may take 
$\gamma =0$ in (\ref{eq:transpulse}). The substitution $s \rightarrow
- i\delta$ transforms (\ref{eq:transpulse}) into a Fourier integral, which
can then be easily evaluated numerically\cite{fisher81,detienne97}.

Fig.~\ref{fig:gauss} shows the result for a gaussian input pulse
${\cal E}_{p}(0,t) = e^{- \alpha t^2}\, e^{i \delta_{0} t}$ which is centered
around frequency $\delta_{0}$ in the spectral domain. In (a) $|{\cal E}_{p}
(L,t)|$ is plotted as a function of $t$ (in units of $L/c$)
for various values of $\delta_{0}$, corresponding to different positions
on the dispersion curve (Fig.~\ref{fig:gauss}(c)).

The vertical line indicates the time $t_{tr}= L/c$ needed to traverse the
mirror in vacuum. We see that an incident pulse which is centered in the 
middle of the gap in the dispersion relation is strongly reshaped upon
transmission and that its peak appears delayed with respect to $t_{tr}$
(curve 1). But if the incoming pulse is centered around a frequency further
away from $\delta = 0$, its peak emerges from the PCM before $t_{tr}$
(curve 2, enlargement in Fig.~\ref{fig:gauss}(b)). The overall reshaping has
become less, but the peak traversal time is superluminal. Moving still further
up the dispersion curve yields a transmitted pulse with shape almost
identical to the incident one and a peak traversal time approaching $t_{tr}$ 
from below (curve 3). This is not surprising, since the frequency components of 
the incident pulse now lie in that part of Fig.~\ref{fig:gauss}(c) 
where the dispersion relation becomes asymptotically linear. The transmitted 
intensity $T_{pp}$ then approaches unity, 
\begin{eqnarray}
T_{pp}(\delta) & \equiv & |t_{pp}(\delta)|^2  = \frac{1 + 
\left(\frac{\delta}{\kappa_{0} c} \right)^2}
{\mbox{\rm cos}^2(\beta L) + \left(\frac{\delta}{\kappa_{0} c} \right)^2}
\\
& \simeq & 1 \hspace{1cm} \mbox{\rm for} \ \ \delta \gg \kappa_{0} c,
\nonumber
\end{eqnarray}
and the pulse propagates almost as in vacuum.

Closer to $\delta=0$, the amplification of the transmitted pulse
is larger and the superluminal peak advancement 
becomes clearly visible, $t_{peak} \simeq 0.75\, t_{tr}$.  
This advancement is a consequence of the increasingly superluminal group 
velocity as $\delta \rightarrow 0$. For a pulse centered around $\delta_{0}$ 
very close to 0,
however, the superluminal effect disappears. Such a pulse contains positive 
as well as negative frequency components, and the presence of
these components with their largely compensating positive and negative group 
velocities leads to strong reshaping and a delayed peak. 
The advancement is thus maximized for pulses centered around a frequency 
$\delta_{0}$ which is small, but still sufficiently far away from 
0 to avoid any influence from the lower half of the dispersion curve.
On the one hand, the pulse should thus be spectrally narrow, in order to be 
as close as possible to the gap region. But on the other hand it should not be
too narrow, since on a temporarily very broad pulse the advancement, 
even if substantial,
would not be easily detectable.
More quantitatively, we have optimized the ratio $\frac{\mbox{\rm 
peak advancement}}{\mbox{\rm pulse width}}$: $r \equiv r(\Delta_{t}, 
\delta_{0}, \kappa_{0}) = \frac{t_{tr} - t_{peak}}
{\Delta_{t}}$, where $\Delta_{t}$ is the temporal width of the
incoming pulse. For the gaussian considered in Fig.~\ref{fig:gauss}
one finds, by varying simultaneously $\Delta_{t}$ and $\delta_{0}$ and
keeping $\kappa_{0}$ fixed, that
the optimal value of $r \simeq 0.073$. For fixed $\delta_{0}$ 
(and $\kappa_{0}$) $r$ decreases fast with increasing 
temporal width of the incoming pulse\cite{pilimit}.

We now demonstrate that the superluminal peak advancement does not 
disagree with the principle of causality, in the sense that the 
peak of the pulse does not transmit any real information.

\noindent To this end, it is convenient to rewrite
(\ref{eq:transpulse}) as
\be
{\cal E}_{p}(L,t) = \int dt^{'} {\cal E}_{p}(0,t^{'}) H(t,t^{'}), 
\label{eq:transpulse2}
\ee
with
\be
 H(t,t^{'}) = \frac{1}{2\pi i} \int_{\gamma - i\infty}^{\gamma
+ i \infty} ds\, t_{pp}(is) e^{s(t- t^{'})} 
\label{eq:Hfunction}
\ee
and $t_{pp}$ given by (\ref{eq:probetrans}). The integral (\ref{eq:Hfunction})
can be evaluated analytically, following an analogous method as used
in\cite{fisher81}. The result is
\be
{\cal E}_{p}(L,t) = \sum_{n=0}^{\infty} \int_{-\infty}^{t - (n+\frac{1}{2}) \tau}
d t^{'} {\cal E}_{p}(0,t^{'})\, L_{n}(t, t^{'}),
\label{eq:Lfunction}
\ee
with
\be
\begin{array}{rl}
L_{n}(t, t^{'}) & = \frac{\kappa_{0} c}{2} A_{n}^{-\frac{1}{2}} 
\left(  
A_{n}^{n} I_{2n -1} \left[ \kappa_{0} c\, \sqrt{(t - t^{'})^2  
- (n+\frac{1}{2})^2 \tau^2}\, \right]  \right.
\vspace{0.2cm} \\
&  - 2 A_{n}^{n+1} I_{2n + 1}\left[ \kappa_{0} c\, \sqrt{(t - t^{'})^2 - 
(n+\frac{1}{2})^2 \tau^2}\, \right]
\vspace{0.2cm} \\
&  \left. + A_{n}^{n+2} I_{2n + 3} \left[ \kappa_{0} c\, \sqrt{(t - t^{'})^2 - 
(n+\frac{1}{2})^2 \tau^2}\, \right]
\ \ \right) \vspace{0.2cm} \\
A_{n} & \equiv \ \frac{t - t^{'} - (n + \frac{1}{2}) \tau}{t - t^{'} + 
(n + \frac{1}{2}) \tau} \vspace{0.3cm} \\
\tau & \equiv  \ 2L/c \ = \ 2\, t_{tr}, \hspace{0.5cm} \mbox{\rm the PCM
roundtrip time.}
\end{array}
\ee
The expression (\ref{eq:Lfunction}) for ${\cal E}_{p}(L,t)$ in terms of 
modified Bessel functions $I_{n}$ is equivalent to the Fourier integral 
and serves as a double check for the results in Fig.~\ref{fig:gauss}.

(\ref{eq:Lfunction}) also gives a good illustration of what happens for an
incoming non-analytic signal, which is suddenly switched on
at $t=0$, so ${\cal E}_{p}(0,t) = {\cal E}_{p}(0,t)\, \Theta(t)$ with
$\Theta$ the Heaviside step function. The transmitted pulse is then
given by\cite{chopped}
\be
\begin{array}{lll}
{\cal E}_{p}(L,t) & = & \sum_{n=0}^{\infty} \Theta(t - (n + \frac{1}{2}) \tau) 
\vspace{0.2cm} \\
& & \ \ \int_{0}^{t - (n+\frac{1}{2}) \tau}
d t^{'} {\cal E}_{p}(0,t^{'}) L_{n}(t, t^{'}).
\end{array}
\label{eq:chop1}
\ee
There is no signal emerging at $x=L$ before time $t= \frac{1}{2} \tau = t_{tr}$,
so the chopped edge of the pulse travels with the speed of light. 
The "information" contained in this abrupt disturbance
is thus transmitted causally. Subsequent
contributions to the sum in (\ref{eq:chop1}) appear after each following 
roundtrip time.

\begin{figure}
\centerline{\epsfig{figure=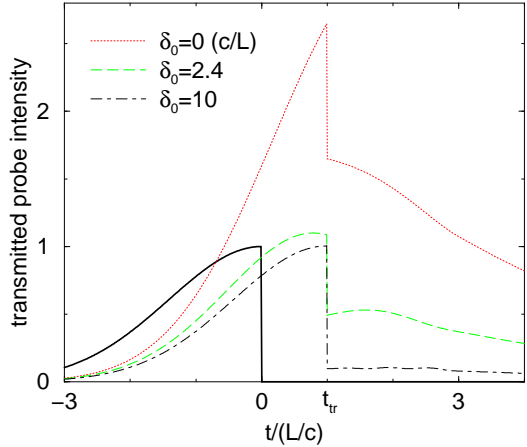,width=0.95\hsize}}
\caption{The transmitted probe pulse through a PCM at $x=L$ for a 
chopped gaussian pulse incident at $x=0$ (thick solid line).
$\kappa_{0} L = 1.4$.
}
\label{fig:choppulse}
\end{figure}

Similar causal transmission is also seen in the response 
at $x=L$ for an incident
gaussian pulse which is suddenly switched off at its maximum, see 
Fig.~\ref{fig:choppulse}.
For all values of $\delta_{0}$ the step is again transmitted causally,
traveling with the speed of light. One clearly sees that for pulses 
centered around intermediate frequencies in the dispersion diagram
($\delta_{0} \simeq 2.4\, c/L$) the previously observed advancement
of the pulse maximum remains, because it is formed before an observer 
at $x=L$ learns about 
the chopped edge of the pulse. After transmission of the step, the 
pulse decays, since there is no further input signal to the mirror
reenforcing and triggering new multiple reflections.

Finally, consider a realistic phase-conjugating mirror, consisting of a cell 
of length $L \sim 10^{-2}$~m. Reflectivities 
on the order of $100\% $, so 
tan$^2(\kappa_{0} L) \sim 1$, have been 
reported for PCM's\cite{lanzerotti96} and hence coupling strengths 
$\kappa_{0} c \sim c/L \sim 10^{10} s^{-1}$ can be reached. As seen 
from the above, a pulse of temporal width
$\sim 0.1$ ns incident on this PCM 
gives rise to peak advance and delay times 
$\sim 0.03$ ns, well within range of observation. 

Summarizing, we have studied the transmission of wave packets 
through a phase-conjugating mirror. Based on the tachyonic dispersion 
relation in the nonlinear PCM medium, superluminal peak traversal times
are predicted.
Thus, a so far unnoticed link is established between optical phase 
conjugation and superluminal behavior, which, especially in view of the
amplifying properties of a PCM,
provides an excellent 
framework for an experimental observation of a signature of optical 
tachyons.

Part of this work was supported by the "Stichting voor Fundamenteel
Onderzoek der Materie" (FOM) which is part of the "Nederlandse Organisatie 
voor Wetenschappelijk Onderzoek" (NWO).
\vspace*{-0.5cm} \\


\begin{references}
\bibitem{validity}  The result (\ref{eq:transpulse}) is valid for arbitrary
		    incident pulses of finite total energy, provided they 
		    decrease faster than exponentially for $t 
		    \rightarrow - \infty$.
\bibitem{pilimit}   The parameter $r$ becomes larger as the coupling constant
		   $\kappa_{0}L$ gets closer to $\pi /2$. Note, however that
		    in the limit $\kappa_{0}L \rightarrow \pi /2$ and $\delta_{0}
		    \rightarrow 0$, when the transmitted probe intensity
		    becomes infinite\cite{fisher83}, our analysis is not 
		     valid anymore, since there pump depletion is bound
		     to reduce the transmitted intensity. A more sophisticated
		     treatment is then required. In this paper we are always 
		     in a situation where pump depletion is negligibly small.
\bibitem{chopped}   Because the Fourier transform of a discontinuous time-pulse
		    contains frequencies ranging from $-\infty$ to $+\infty$,
		    care has to be taken in applying (\ref{eq:transpulse}) 
		    to these kind of pulses, since (\ref{eq:transpulse}) is 
		    based on the assumption $\delta \ll \omega_{0}$. However,
		    for the chopped pulses considered here, the weight of
		    frequency components close to $\omega_{0}$ (which is 
		    typically $\sim 10^{15} s^{-1}$, while $\delta 
		    \sim 10^8 s^{-1}$) 
		    is negligibly small.
\end{references}
\end{document}